\documentclass[11pt, a4paper]{article}
\usepackage[utf8]{inputenc}
\usepackage{amsmath,setspace,geometry}
\usepackage{amsthm}
\usepackage{amsfonts}
\usepackage[shortlabels]{enumitem}
\usepackage{rotating}
\usepackage{pdflscape}
\usepackage{graphicx}
\usepackage{bbm}
\usepackage[dvipsnames]{xcolor}
\usepackage{hyperref}
\hypersetup{colorlinks=true, linkcolor= BrickRed, citecolor = BrickRed, filecolor = BrickRed, urlcolor = BrickRed, hypertexnames = true}
\usepackage[]{natbib} 
\bibpunct[:]{(}{)}{,}{a}{}{,}
\usepackage[english]{babel}
\usepackage{float}
\usepackage{caption}
\usepackage{subcaption}
\usepackage{booktabs}
\usepackage{pdfpages}
\usepackage{threeparttable}
\usepackage{lscape}
\usepackage{bm}
\newtheorem{proposition}{Proposition}

\title{Resolving the Conflict on Conduct Parameter Estimation in Homogeneous Goods Markets between Bresnahan (1982) and Perloff and Shen (2012)}
\author{Yuri Matsumura\thanks{Department of Economics, Rice University. Email: Yuri.Matsumura@rice.edu} \and Suguru Otani \thanks{Department of Economics, Rice University. Email: so19@rice.edu
}}

\begin{document}

\maketitle
\begin{abstract}
    We revisit conduct parameter estimation in homogeneous goods markets to resolve the conflict between Bresnahan (1982) and Perloff and Shen (2012) regarding the identification and the estimation of conduct parameters. We point out that Perloff and Shen's (2012) proof is incorrect and its simulation setting is invalid. Our simulation shows that estimation becomes accurate when demand shifters are properly added in supply estimation and sample sizes are increased, supporting Bresnahan (1982).
\vspace{0.1in}

\vspace{0in}

\bigskip
\end{abstract}

\section{Introduction}
Measuring competitiveness is an important task in the empirical industrial organization literature.
A conduct parameter is considered to be a useful measure of competitiveness. 
However, the parameter cannot be directly measured from data because data generally lack information about marginal costs.
Therefore, researchers endeavor to identify and estimate conduct parameters.

In this regard, there are two conflicting results regarding conduct parameter estimation in homogeneous goods markets with linear demand and marginal cost systems.
On the one hand, \citet{bresnahan1982oligopoly} proposes an approach to identify a conduct parameter using demand rotation instruments.
With identification guaranteed, the conduct parameter can be estimated using standard linear regression.
This result is extended to nonlinear cases by \citet{lau1982identifying} and differentiated product markets by \citet{nevoIdentificationOligopolySolution1998}.

On the other hand, \citet{perloff2012collinearity} (hereafter, PS) assert that the linear model considered by \citet{bresnahan1982oligopoly} suffers from a multicollinearity problem when error terms in the demand and supply equations are zero, implying that identification of the conduct parameter is impossible.
Moreover, PS used simulations to demonstrate that the conduct parameter cannot be estimated accurately even when the error terms are nonzero. 
This disagreement is a major obstacle in the literature. 
Several papers and handbook chapters have referenced PS’s results, such as \citet{claessensWhatDrivesBank2004, coccoreseMultimarketContactCompetition2013, coccoreseWhatAffectsBank2021, garciaMarketStructuresProduction2020, kumbhakarNewMethodEstimating2012, perekhozhukRegionalLevelAnalysisOligopsony2015}, and \citet{shafferMarketPowerCompetition2017}.

We revisit conduct parameter identification and estimation in homogeneous product markets to determine the validity of these results.
First, we show that the proof of the multicollinearity problem in PS is incorrect and that the problem does not occur under standard assumptions reflecting the insights by \citet{bresnahan1982oligopoly}.
Second, the simulations in PS lack an excluded demand shifter in supply estimation; we confirm that estimation is accurate when including a demand shifter in supply estimation. 
We also show that increasing sample size improves accuracy of estimation. 
Hence, our results support those of \cite{bresnahan1982oligopoly} theoretically and numerically.

\section{Model}
Consider data with $T$ markets with homogeneous products.
Assume that there are $N$ firms in each market.
Let $t = 1,\ldots, T$ be the index for markets.
Then, we obtain a supply equation as follows:
\begin{align}
     P_t = -\theta\frac{\partial P_t(Q_{t})}{\partial Q_{t}}Q_{t} + MC_t(Q_{t}),\label{eq:supply_equation}
\end{align}
where $Q_{t}$ is the aggregate quantity, $P_t(Q_{t})$ is the demand function, $MC_{t}(Q_{t})$ is the marginal cost function, and $\theta\in[0,1]$ is  the conduct parameter. 
The equation nests perfect competition ($\theta=0$), Cournot competition ($\theta=1/N$), and perfect collusion ($\theta=1$).\footnote{See \cite{bresnahan1982oligopoly}.} 

Consider an econometric model that integrates the above model.
Assume that the demand and marginal cost functions are written as follows: 
\begin{align}
    P_t = f(Q_{t}, Y_t, \varepsilon^{d}_{t}, \alpha), \label{eq:demand}\\
    MC_t = g(Q_{t}, W_{t}, \varepsilon^{c}_{t}, \gamma),\label{eq:marginal_cost}
\end{align}
where $Y_t$ and $W_{t}$ are vectors of exogenous variables, $\varepsilon^{d}_{t}$ and $\varepsilon^{c}_{t}$ are error terms, and $\alpha$ and $\gamma$ are vectors of parameters.
Additionally, we have demand- and supply-side instrumental variables, $Z^{d}_{t}$ and $Z^{c}_{t}$, and assume that the error terms satisfy the mean independence conditions, $E[\varepsilon^{d}_{t}\mid Y_t, Z^{d}_{t}] = E[\varepsilon^{c}_{t} \mid W_{t}, Z^{c}_{t}] =0$.

\subsection{Linear demand and cost}
Assume that linear demand and marginal cost functions are specified as follows:
\begin{align}
    P_t &= \alpha_0 - (\alpha_1 + \alpha_2Z^{R}_{t})Q_{t} + \alpha_3 Y_t + \varepsilon^{d}_{t},\label{eq:linear_demand}\\
    MC_t &= \gamma_0  + \gamma_1 Q_{t} + \gamma_2 W_{t} + \gamma_3 R_{t} + \varepsilon^{c}_{t},\label{eq:linear_marginal_cost}
\end{align}
where $W_{t}$ and $R_{t}$ are excluded cost shifters and $Z^{R}_{t}$ is Bresnahan's demand rotation instrument. 
The supply equation is written as follows:
\begin{align}
    P_t 
    &= \gamma_0 + \theta \alpha_2 Z^{R}_tQ_{t} + (\theta\alpha_1 + \gamma_1) Q_{t} + \gamma_2 W_t + \gamma_3 R_{t} +\varepsilon^c_t.\label{eq:linear_supply_equation}
\end{align}
By substituting Equation \eqref{eq:linear_demand} with Equation \eqref{eq:linear_supply_equation} and solving it for $P_t$, we obtain the aggregate quantity $Q_{t}$ based on the parameters and exogenous variables as follows:
\begin{align}
    Q_{t} =  \frac{\alpha_0 + \alpha_3 Y_t - \gamma_0 - \gamma_2 W_{t} - \gamma_3 R_{t} + \varepsilon^{d}_{t} - \varepsilon^{c}_{t}}{(1 + \theta) (\alpha_1 + \alpha_2 Z^{R}_{t}) + \gamma_1}.\label{eq:quantity_linear}
\end{align}

\subsection{Is the multicollinearity problem in PS incorrect?}
To demonstrate the multicollinearity problem, PS attempt to demonstrate linear dependence between the variables in the supply equation. 
PS begin the proof on page 137 in their appendix by stating the following (we modify the notation):
\begin{quote}
    ``We demonstrate that the $W_{t}, R_{t}, Z^{R}_{t}Q_{t}$, and $Q_{t}$ terms in Eq.4 are perfectly collinear for $\varepsilon_{t}^{d} = \varepsilon_{t}^{c} = 0$. We show this result by demonstrating that there exist nonzero coefficients $\chi_1,\chi_2,\chi_3,\chi_4$, and $\chi_5$ such that 
   \begin{align*}
    Z^{R}_{t} Q_{t} + \chi_1 Q_{t} + \chi_2 W_{t} + \chi_3 R_{t} + \chi_4 Y_{t} + \chi_5 = 0 \quad (\text{A1})."
    \end{align*}
\end{quote}
Eq.4 in the quotation corresponds to the supply equation \eqref{eq:linear_supply_equation}.
Therefore, PS show that there exists a nonzero vector of $\chi_1, \ldots, \chi_5$ that satisfies (A1).

An incorrect detail in the proof is that while attempting to demonstrate linear dependence between $Z^{R}_{t}Q_{t}, Q_{t}, W_{t}$, and $R_{t}$, they show linear dependence between $Z^{R}_{t}Q_{t}, Q_{t}, W_{t}, R_{t}$, and $Y_t$. 
However, linear dependence between $Z^{R}_{t}Q_{t}, Q_{t},W_{t}, R_{t}$, and $Y_t$ does not always imply linear dependence between $Z^{R}_{t}Q_{t}, Q_{t}, W_{t}$, and $R_{t}$.

Therefore, we contend that a multicollinearity problem does not occur under the additional standard assumptions in Proposition 1.
\begin{proposition}
    Assume that (i) $\alpha_2$ and $\alpha_3$ are nonzero and (ii) $Z^R_t, W_t, R_t$, and $Y_t$ are linearly independent.
    Then, $Z^{R}_{t}Q_{t}, Q_{t}, W_{t}$, and $R_{t}$ are linearly independent.
\end{proposition}

See online appendix for the proof.
Equation \eqref{eq:linear_supply_equation} implies that the main challenge is separately identifying the conduct parameter and the slope of marginal cost.
As quantity is endogenous, this requires two excluded instruments. 
The assumption makes the demand rotation instrument and the demand shifter relevant and ensures that these instruments and the other cost shifters do not covary.
Under the assumption, identification of the conduct parameter is possible.

In the context of differentiated products markets, \cite{magnolfi2022falsifying} discuss similar issues concerning instrument requirements for falsifying models with upward sloping marginal cost. 
They build on the results of \cite{berry2014identification}, who show that with instruments, falsification of models of conduct with flexible cost functions is possible.

\section{Simulation results}\label{sec:results}

Table \ref{tb:linear_linear_sigma_1} presents the results of estimating the linear model with the demand shifter.\footnote{See online appendix for simulation details and additional results.}
Panel (a) shows that when the standard deviations of the error terms in the demand and supply equations are $\sigma = 0.001$, estimation of all parameters is extremely accurate.
When sample size is large, root-mean-squared errors (RMSEs) of all parameters are less than or equal to 0.001. 
Panel (c) shows the case with $\sigma = 2.0$. 
As sample size increases, the RMSEs sharply decrease. 
Thus, the imprecise results reported by PS are due to the lack of demand shifters and small sample size.

\begin{table}[!htbp]
  \begin{center}
      \caption{Results of the linear model with demand shifter}
      \label{tb:linear_linear_sigma_1} 
      \subfloat[$\sigma=0.001$]{\begin{tabular}[t]{llrrrrrrr}
\toprule
  & Bias & RMSE & Bias & RMSE & Bias & RMSE & Bias & RMSE\\
\midrule
$\alpha_{0}$ & 0.000 & 0.001 & 0.000 & 0.001 & 0.000 & 0.000 & 0.000 & 0.000\\
$\alpha_{1}$ & 0.000 & 0.004 & 0.000 & 0.003 & 0.000 & 0.002 & 0.000 & 0.001\\
$\alpha_{2}$ & 0.000 & 0.000 & 0.000 & 0.000 & 0.000 & 0.000 & 0.000 & 0.000\\
$\alpha_{3}$ & 0.000 & 0.000 & 0.000 & 0.000 & 0.000 & 0.000 & 0.000 & 0.000\\
$\gamma_{0}$ & 0.000 & 0.001 & 0.000 & 0.001 & 0.000 & 0.001 & 0.000 & 0.000\\
$\gamma_{1}$ & 0.000 & 0.005 & 0.000 & 0.004 & 0.000 & 0.002 & 0.000 & 0.001\\
$\gamma_{2}$ & 0.000 & 0.000 & 0.000 & 0.000 & 0.000 & 0.000 & 0.000 & 0.000\\
$\gamma_{3}$ & 0.000 & 0.000 & 0.000 & 0.000 & 0.000 & 0.000 & 0.000 & 0.000\\
$\theta$ & 0.000 & 0.001 & 0.000 & 0.000 & 0.000 & 0.000 & 0.000 & 0.000\\
Sample size ($T$) &  & 50 &  & 100 &  & 200 &  & 1000\\
\bottomrule
\end{tabular}}\\
      \subfloat[$\sigma=0.5$]{\begin{tabular}[t]{llrrrrrrr}
\toprule
  & Bias & RMSE & Bias & RMSE & Bias & RMSE & Bias & RMSE\\
\midrule
$\alpha_{0}$ & -0.018 & 0.465 & 0.007 & 0.323 & -0.008 & 0.213 & -0.006 & 0.097\\
$\alpha_{1}$ & -0.045 & 2.257 & 0.024 & 1.523 & 0.018 & 1.016 & -0.031 & 0.455\\
$\alpha_{2}$ & -0.001 & 0.255 & -0.001 & 0.176 & -0.004 & 0.115 & 0.001 & 0.051\\
$\alpha_{3}$ & -0.005 & 0.108 & 0.003 & 0.075 & -0.001 & 0.050 & -0.001 & 0.022\\
$\gamma_{0}$ & -0.061 & 0.732 & -0.005 & 0.474 & -0.021 & 0.346 & -0.005 & 0.152\\
$\gamma_{1}$ & -0.311 & 3.450 & -0.124 & 1.928 & -0.081 & 1.303 & -0.003 & 0.548\\
$\gamma_{2}$ & 0.009 & 0.109 & -0.001 & 0.071 & 0.003 & 0.051 & 0.000 & 0.023\\
$\gamma_{3}$ & 0.001 & 0.108 & 0.003 & 0.075 & 0.003 & 0.053 & 0.000 & 0.022\\
$\theta$ & 0.047 & 0.354 & 0.017 & 0.209 & 0.014 & 0.135 & 0.003 & 0.058\\
Sample size ($T$) &  & 50 &  & 100 &  & 200 &  & 1000\\
\bottomrule
\end{tabular}}\\
    \subfloat[$\sigma=2.0$]{\begin{tabular}[t]{llrrrrrrr}
\toprule
  & Bias & RMSE & Bias & RMSE & Bias & RMSE & Bias & RMSE\\
\midrule
$\alpha_{0}$ & -0.263 & 2.596 & 0.071 & 1.670 & -0.040 & 0.947 & -0.002 & 0.412\\
$\alpha_{1}$ & -0.271 & 10.820 & 0.008 & 6.492 & 0.236 & 4.263 & 0.021 & 1.809\\
$\alpha_{2}$ & -0.044 & 1.253 & 0.023 & 0.779 & -0.031 & 0.483 & -0.003 & 0.210\\
$\alpha_{3}$ & -0.024 & 0.584 & 0.008 & 0.343 & -0.004 & 0.225 & 0.003 & 0.092\\
$\gamma_{0}$ & -2.074 & 19.624 & -0.551 & 3.043 & -0.171 & 1.516 & -0.051 & 0.633\\
$\gamma_{1}$ & 58.209 & 1750.688 & -2.416 & 56.909 & -3.617 & 39.044 & -0.103 & 2.334\\
$\gamma_{2}$ & 0.242 & 2.430 & 0.065 & 0.409 & 0.020 & 0.220 & 0.006 & 0.093\\
$\gamma_{3}$ & 0.230 & 2.328 & 0.055 & 0.404 & 0.010 & 0.219 & 0.008 & 0.092\\
$\theta$ & -6.668 & 233.851 & 0.372 & 6.334 & 0.418 & 3.820 & 0.024 & 0.245\\
Sample size ($T$) &  & 50 &  & 100 &  & 200 &  & 1000\\
\bottomrule
\end{tabular}}
  \end{center}
  \footnotesize
  Note: The error terms in the demand and supply equation are drawn from a normal distribution, $N(0,\sigma)$.
\end{table}

\section{Conclusion}
We revisit conduct parameter estimation in homogeneous goods markets.
There is a conflict between \citet{bresnahan1982oligopoly} and \citet{perloff2012collinearity} in terms of identification and estimation.
We highlight problems in the proof and the simulation in \citet{perloff2012collinearity}.
Our simulation shows that estimation of the conduct parameter becomes accurate when demand shifters are appropriately introduced in supply estimation and sample size is increased. 
Based on our theoretical and numerical investigation, we support the argument made by \citet{bresnahan1982oligopoly}.

\paragraph{Acknowledgments}
We thank Jeremy Fox and Isabelle Perrigne for their valuable advice. This research did not receive any specific grant from funding agencies in the public, commercial, or not-for-profit sectors. 

\newpage

\bibliographystyle{aer}
\bibliography{conduct_parameter}

\newpage

\setcounter{page}{1}
\appendix
\section{Online appendix}\label{sec:appendix}
\subsection{Omitted proof of Proposition 1}
\begin{proof}
    Based on the definition of linear independence, we need to confirm that the following holds:
\begin{align}
    \chi_1 Z_{t}^R Q + \chi_2 Q_{t} + \chi_3 W_{t} + \chi_4 R_{t} + \chi_5 = 0, \label{eq:linear_independence}
\end{align}
then $\chi_1 = \chi_2 = \cdots = \chi_5 = 0$.

By substituting Equation \eqref{eq:quantity_linear} into Equation \eqref{eq:linear_independence}, we obtain the following:
\begin{align*}
    0 &= \zeta_1 Z_{t}^R + \zeta_2 Z_{t}^RY_{t} + \zeta_3 W_{t}Z_{t}^R + \zeta_4 R_{t}Z_{t}^R + \zeta_5 Y_{t} + \zeta_6 W_{t} + \zeta_7 R_{t} + \zeta_8, 
\end{align*}
where 
\begin{align*}
    \zeta_1 &= (\alpha_0 - \gamma_0)\chi_1  + (\theta +1 )\alpha_2 \chi_5 ,\\
    \zeta_2 &= \alpha_3\chi_1,\\
    \zeta_3 &= -\gamma_2 \chi_1 + (\theta + 1)\alpha_2\chi_3,\\
    \zeta_4 &= -\gamma_3 \chi_1 + (\theta + 1)\alpha_2\chi_4,\\
    \zeta_5 &=  \alpha_3\chi_2,\\
    \zeta_6 &= -\gamma_2 \chi_2 + [(1 + \theta) \alpha_1 +\gamma_1]\chi_3,\\
    \zeta_7 &= -\gamma_3 \chi_2 +  [(1 + \theta) \alpha_1 +\gamma_1]\chi_4,\\
    \zeta_8 &=  (\alpha_0 - \gamma_0)\chi_2 +[(1 + \theta)\alpha_1 +\gamma_1] \chi_5.
\end{align*}

First, based on Assumption (ii), $\zeta_1 = \cdots = \zeta_8 = 0$.
Second, as the parameters are nonzero by Assumption (i), $\chi_1 = \chi_2 =0$ by $\zeta_2 = \zeta_5 = 0$.
Third, by $\zeta_1 = \zeta_3 = \zeta_4 = 0$, $(\theta + 1 )\alpha_2\chi_5 = (\theta + 1 )\alpha_2\chi_3 = (\theta + 1 )\alpha_2\chi_4 = 0.$
As $(\theta + 1)\alpha_2 \ne 0$ by Assumption (i), $\chi_3 = \chi_4 = \chi_5 = 0$.
This completes the proof.
\end{proof}

\subsection{Simulation and estimation procedure}

We set true parameters and distributions as shown in Table \ref{tb:parameter_setting}. 
We follow the setting of PS. For simulation, we generate 1,000 data sets.
We separately estimate the demand and supply equation by using two-stage least squares (2SLS) estimation.
The instrumental variables for demand estimation are $Z^{d}_{t} = (Z^{R}_{t}, Y_t, H_{t}, K_{t})$ and the instrumental variables for supply estimation are $Z^{c}_{t} = (Z^{R}_{t}, W_{t}, R_{t}, Y_t)$. 

\begin{table}[!htbp]
    \caption{True parameters and distributions}
    \label{tb:parameter_setting}
    \begin{center}
    \subfloat[Parameters]{
    \begin{tabular}{cr}
            \hline
            & linear  \\
            $\alpha_0$ & $10.0$  \\
            $\alpha_1$ & $1.0$  \\
            $\alpha_2$ & $1.0$ \\
            $\alpha_3$ & $1.0$  \\
            $\gamma_0$ & $1.0$ \\
            $\gamma_1$ & $1.0$  \\
            $\gamma_2$ & $1.0$ \\
            $\gamma_3$ & $1.0$\\
            $\theta$ & $0.5$ \\
            \hline
        \end{tabular}
    }
    \subfloat[Distributions]{
    \begin{tabular}{crr}
            \hline
            & linear\\
            Demand shifter&  \\
            $Y_t$ & $N(0,1)$  \\
            Demand rotation instrument&   \\
            $Z^{R}_{t}$ & $N(10,1)$ \\
            Cost shifter&    \\
            $W_{t}$ & $N(3,1)$  \\
            $R_{t}$ & $N(0,1)$   \\
            $H_{t}$ & $W_{t}+N(0,1)$  \\
            $K_{t}$ & $R_{t}+N(0,1)$   \\
            Error&  &  \\
            $\varepsilon^{d}_{t}$ & $N(0,\sigma)$  \\
            $\varepsilon^{c}_{t}$ & $N(0,\sigma)$ \\
            \hline
        \end{tabular}
    }
    \end{center}
    \footnotesize
    Note: $\sigma=\{0.001, 0.5, 2.0\}$. $N:$ Normal distribution. $U:$ Uniform distribution.
\end{table}

\subsection{Details for our simulation settings}

To generate the simulation data, for each model, we first generate the exogenous variables $Y_t, Z^{R}_{t}, W_t, R_{t}, H_t$, and $K_t$ and the error terms $\varepsilon_{t}^c$ and $\varepsilon_{t}^d$ based on the data generation process in Table \ref{tb:parameter_setting}.
We compute the equilibrium quantity $Q_{t}$ for the linear model by \eqref{eq:quantity_linear}.
We then compute the equilibrium price $P_t$ by substituting $Q_{t}$ and other variables into the demand function \eqref{eq:linear_demand}.

We estimate the equations using the \texttt{ivreg} package in \texttt{R}.
An important feature of the model is that we have an interaction term of the endogenous variable $Q_{t}$ and the instrumental variable $Z^{R}_{t}$.
The \texttt{ivreg} package automatically detects that the endogenous variables are $Q_{t}$ and the interaction term $Z^{R}_{t}Q_{t}$, running the first stage regression for each endogenous variable with the same instruments. To confirm this, we manually write R code to implement the 2SLS model. 
When the first stage includes only the regression of $Q_{t}$, estimation results from our code differ from the results from \texttt{ivreg}. 
However, when we modify the code to regress $Z^{R}_{t}Q_{t}$ on the instrument variables and estimate the second stage by using the predicted values of $Q_{t}$ and $Z^{R}_{t}Q_{t}$, the result from our code and the result from \texttt{ivreg} coincide.

\subsection{Other experiments}

\begin{table}[!htbp]
    \caption{Estimation results in Table 2 of from PS}
    \label{tb:linear_linear_sigma_Perloff_Shen}
    \begin{center}
        \begin{tabular}{cllll}
            \hline
            & $\sigma=0.001$ & $\sigma=0.5$ & $\sigma=1$ & $\sigma=2$ \\
            $\alpha_0$ & $10.00\ (0.001)$ & $9.96\ (0.33)$ & $9.86\ (0.65)$ & $9.46 (1.20)$ \\
            $\alpha_1$ & $1.00\ (0.004)$ & $0.99\ (1.98)$ & $0.97\ (3.96)$ & $0.88 (7.80)$ \\
            $\alpha_2$ & $1.00\ (0.004)$ & $0.99\ (0.21)$ & $0.97\ (0.42)$ & $0.87\ (0.82)$ \\
            $\gamma_1$ & $0.46\ (0.88)$ & $0.46\ (0.91)$ & $0.47\ (0.93)$ & $0.49\ (1.04)$ \\
            $\gamma_2$ & $5.85\ (7.89)$ & $5.85\ (8.15)$ & $5.78\ (8.21)$ & $5.73\ (8.66)$ \\
            $\theta$ & $-0.31\ (1.31)$ & $-0.29\ (1.34)$ & $0.09\ (11.48)$ & $-1.53\ (30.41)$ \\
            \hline
        \end{tabular}
    \end{center}\footnotesize
    Note: True parameters: $\alpha_1 = \alpha_2 = \gamma_0 = \gamma_1 = \gamma_2  = \gamma_3 = 1, \alpha_0 = 10, \alpha_3 = 0,$ and $\theta = 0.5$. PS exclude $Y_t$. We change the parameter notations from the original study. Note that PS do not provide $\gamma_0$ and $\gamma_3$.
\end{table}

First, we replicate the result in PS. For comparison, we report the means and standard deviations (SDs).
To replicate the result, we exclude the demand shifter $Y_t$ and assume the coefficient $\alpha_3$ of $Y_t$ is zero, indicating that there is no demand shifter for supply estimation.
For reference, Table \ref{tb:linear_linear_sigma_Perloff_Shen} is quoted from PS, although we modify some notation.
Sample size in each simulation dataset is 50 and the table shows the means and SDs of the 2SLS estimators from 1,000 simulations.
It shows that demand estimation becomes more accurate as the values of the SDs of the error terms, that is, $\sigma$ decreases.
In contrast, supply estimation is still biased and the SD of the conduct parameter becomes larger as the value of $\sigma$ increases.

Our replication results are presented in Table \ref{tb:linear_linear_sigma_1_without_demand_shifter_y}.
Each panel presents the simulation results under different SDs.
This result uses the same data generation process as PS. 
To determine whether we can correctly replicate the result in PS, we focus on the first two columns in each panel.
These two columns show the means and SDs of the simulation results when sample size is 50.
While demand parameters can be accurately estimated, although the value of $\sigma$ becomes higher, the supply parameters are biased.
In particular, when $\sigma$ is large and sample size is small, the SDs of the parameters in supply equation become large.
Thus, we reveal the patterns in PS that do not provide any details.

As PS fix sample size to 50, we also examine the effect of changing sample size.
As expected, increasing sample size given a value of $\sigma$ decreases the SDs of supply parameters.
However, no simulation result is close to the true values of supply parameters as well as the conduct parameter.
These results are consistent with PS.

\begin{table}[!htbp]
  \begin{center}
      \caption{Estimation results of the linear model without demand shifter}
      \label{tb:linear_linear_sigma_1_without_demand_shifter_y} 
      \subfloat[$\sigma=0.001$]{\begin{tabular}[t]{lrrrrrrrr}
\toprule
  & Mean & SD & Mean  & SD  & Mean   & SD   & Mean    & SD   \\
\midrule
$\alpha_{0}$ & 10.000 & 0.001 & 10.000 & 0.001 & 10.000 & 0.000 & 10.000 & 0.000\\
$\alpha_{1}$ & 1.000 & 0.004 & 1.000 & 0.003 & 1.000 & 0.002 & 1.000 & 0.001\\
$\alpha_{2}$ & 1.000 & 0.000 & 1.000 & 0.000 & 1.000 & 0.000 & 1.000 & 0.000\\
$\gamma_{0}$ & 5.446 & 6.981 & 5.388 & 7.986 & 5.423 & 7.825 & 5.063 & 6.801\\
$\gamma_{1}$ & 0.506 & 0.775 & 0.512 & 0.888 & 0.509 & 0.869 & 0.549 & 0.756\\
$\gamma_{2}$ & 0.506 & 0.776 & 0.512 & 0.887 & 0.509 & 0.869 & 0.549 & 0.756\\
$\theta$ & -0.241 & 1.164 & -0.231 & 1.331 & -0.237 & 1.304 & -0.177 & 1.134\\
$R^{2}$ (demand) & 1.000 & 0.000 & 1.000 & 0.000 & 1.000 & 0.000 & 1.000 & 0.000\\
$R^{2}$ (supply) & 1.000 & 0.000 & 1.000 & 0.000 & 1.000 & 0.000 & 1.000 & 0.000\\
Sample size ($T$) &  & 50 &  & 100 &  & 200 &  & 1000\\
\bottomrule
\end{tabular}}\\
      \subfloat[$\sigma=0.5$]{\begin{tabular}[t]{lrrrrrrrr}
\toprule
  & Mean & SD & Mean  & SD  & Mean   & SD   & Mean    & SD   \\
\midrule
$\alpha_{0}$ & 9.993 & 0.466 & 9.993 & 0.312 & 10.001 & 0.215 & 10.002 & 0.093\\
$\alpha_{1}$ & 0.963 & 2.138 & 0.965 & 1.484 & 1.012 & 1.023 & 0.991 & 0.441\\
$\alpha_{2}$ & 1.002 & 0.243 & 1.002 & 0.168 & 0.999 & 0.118 & 1.002 & 0.049\\
$\gamma_{0}$ & 5.332 & 10.459 & 5.227 & 11.592 & 5.112 & 15.871 & 5.470 & 7.476\\
$\gamma_{1}$ & 0.405 & 3.214 & 0.434 & 1.989 & 0.474 & 1.744 & 0.516 & 1.102\\
$\gamma_{2}$ & 0.517 & 1.157 & 0.528 & 1.222 & 0.546 & 1.816 & 0.504 & 0.830\\
$\theta$ & -0.210 & 1.879 & -0.206 & 1.951 & -0.186 & 2.705 & -0.247 & 1.238\\
$R^{2}$ (demand) & 0.720 & 0.088 & 0.725 & 0.061 & 0.726 & 0.041 & 0.728 & 0.018\\
$R^{2}$ (supply) & 0.160 & 7.674 & -0.119 & 19.529 & -0.724 & 30.775 & 0.491 & 2.041\\
Sample size ($T$) &  & 50 &  & 100 &  & 200 &  & 1000\\
\bottomrule
\end{tabular}}\\
  \end{center}\footnotesize
  Note: True parameters: $\alpha_1 = \alpha_2 =  \gamma_0 = \gamma_1 = \gamma_2  =  1, \alpha_0 = 10, \theta = 0.5.$ and $\alpha_3 =0$. For comparison, we report the mean and SD.
\end{table} 

\begin{table}[!htbp]
  \ContinuedFloat
  \begin{center}
      \caption{Estimation results of the linear model without demand shifter (Continued)}
      \subfloat[$\sigma=1.0$]{\begin{tabular}[t]{lrrrrrrrr}
\toprule
  & Mean & SD & Mean  & SD  & Mean   & SD   & Mean    & SD   \\
\midrule
$\alpha_{0}$ & 9.975 & 0.964 & 9.953 & 0.636 & 10.007 & 0.441 & 9.991 & 0.189\\
$\alpha_{1}$ & 1.120 & 4.491 & 0.942 & 2.885 & 0.883 & 2.055 & 1.035 & 0.902\\
$\alpha_{2}$ & 0.981 & 0.492 & 0.993 & 0.326 & 1.015 & 0.227 & 0.993 & 0.101\\
$\gamma_{0}$ & 5.631 & 9.410 & 5.520 & 7.580 & 5.161 & 9.226 & 5.556 & 7.424\\
$\gamma_{1}$ & -0.107 & 19.285 & 0.129 & 5.240 & 0.488 & 3.541 & 0.489 & 1.210\\
$\gamma_{2}$ & 0.476 & 1.043 & 0.495 & 0.835 & 0.540 & 1.030 & 0.494 & 0.820\\
$\theta$ & -0.201 & 3.603 & -0.217 & 1.478 & -0.183 & 1.528 & -0.260 & 1.229\\
$R^{2}$ (demand) & 0.205 & 0.357 & 0.234 & 0.221 & 0.232 & 0.150 & 0.245 & 0.060\\
$R^{2}$ (supply) & -0.920 & 17.898 & -0.395 & 5.271 & -0.904 & 12.486 & -0.421 & 12.047\\
Sample size ($T$) &  & 50 &  & 100 &  & 200 &  & 1000\\
\bottomrule
\end{tabular}}\\
    \subfloat[$\sigma=2.0$]{\begin{tabular}[t]{lrrrrrrrr}
\toprule
  & Mean & SD & Mean  & SD  & Mean   & SD   & Mean    & SD   \\
\midrule
$\alpha_{0}$ & 9.515 & 6.752 & 9.912 & 1.479 & 9.987 & 0.943 & 9.987 & 0.396\\
$\alpha_{1}$ & 0.362 & 19.344 & 0.710 & 6.192 & 1.154 & 4.363 & 0.986 & 1.728\\
$\alpha_{2}$ & 0.934 & 1.092 & 1.004 & 0.743 & 0.981 & 0.494 & 0.998 & 0.204\\
$\gamma_{0}$ & 5.658 & 6.892 & 5.464 & 8.387 & 5.695 & 8.243 & 5.572 & 10.796\\
$\gamma_{1}$ & 0.956 & 52.166 & 1.715 & 42.062 & -0.056 & 11.467 & 0.388 & 3.140\\
$\gamma_{2}$ & 0.479 & 0.827 & 0.496 & 0.907 & 0.486 & 0.902 & 0.497 & 1.185\\
$\theta$ & -0.296 & 5.941 & -0.439 & 5.106 & -0.235 & 2.034 & -0.256 & 1.771\\
$R^{2}$ (demand) & -3.456 & 87.362 & -0.513 & 1.557 & -0.436 & 0.563 & -0.376 & 0.185\\
$R^{2}$ (supply) & -1.104 & 5.881 & -2.311 & 26.606 & -1.993 & 26.973 & -3.591 & 49.060\\
Sample size ($T$) &  & 50 &  & 100 &  & 200 &  & 1000\\
\bottomrule
\end{tabular}}
  \end{center}\footnotesize
  Note: True parameters: $\alpha_1 = \alpha_2 =  \gamma_0 = \gamma_1 = \gamma_2  =  1, \alpha_0 = 10, \theta = 0.5.$ and $\alpha_3 =0$. For comparison, we report the mean and SD.
\end{table}

\end{document}